\documentstyle[12pt]{article}
\input math_macros.tex

\def\ref#1{$^{#1)}$}
\begin{document}
\begin{titlepage}
\begin{center}
June 26, 1992     \hfill    LBL-32531\\

\vskip .5in

{\large \bf Global Quantization in Gauge Orbit Space with Magnetic Monopoles As
a Solution to Strong CP Problem and the Relevance to $U_A(1)$ Problem}
\footnote{This work was supported by the Director, Office of Energy
Research, Office of High Energy and Nuclear Physics, Division of High
Energy Physics of the U.S. Department of Energy under Contract
DE-AC03-76SF00098.}
\vskip .5in

Huazhong Zhang\\[.5in]
{\em Theoretical Physics Group\\
     Lawrence Berkeley Laboratory\\
     MS 50A-3115, 1 Cyclotron Road\\
     Berkeley, California 94720}
\end{center}

\end{titlepage}
\renewcommand{\thepage}{\roman{page}}
\setcounter{page}{2}
\mbox{ }

\vskip 1in

\begin{center}
{\bf Disclaimer}
\end{center}

\vskip .2in

\newpage
\renewcommand{\thepage}{\arabic{page}}
\setcounter{page}{1}
\begin{abstract}
We generalize our discussions and give more general physical applications of a
new solution to the strong CP problem with magnetic monopoles as originally
proposed by the author$^1$. Especially, we will discuss about the global
topological structure in the relevant gauge orbit spaces to be clarified. As it
is shown that in non-abelian gauge theories with a $\theta$ term, the induced
gauge orbit space with gauge potentials and gauge functions restricted on the
space boundary $S^2$ has a magnetic monopole structure and
the gauge orbit space has a vortex structure if there is a magnetic monopole in
the ordinary space. The Dirac's quantization conditions in the quantum theories
ensure that the vacuum angle $\theta$ in the gauge theories must be quantized.
The quantization rule is given by $\theta=2\pi/n~(n\neq 0)$ with n being the
topological charge of the magnetic monopole. Therefore, the strong CP problem
is automatically solved in the presence of a magnetic monopole of charge
$\pm 1$ with $\theta=\pm 2\pi$, or magnetic monopoles of very large total
topological charge ($|n|\geq 10^92\pi$) if it is consistent with the abundance
of magnetic monopoles. Where in the first case with a magnetic monopole of
topological charge 1 or -1, we mean the strong CP-violation can be only very
small by the measurements implemented so far. Since $\theta=\pm 2\pi$
correspond to different monopole sectors, the CP can not be conserved exactly
in strong interactions in this case. In the second case, the strong CP cannot
be conserved either for large but finite n. The fact that the strong
CP-violation measured so far can be only so small or vanishing may be a signal
for the existence of magnetic monopoles. We also conjecture that the parity
violation and CP violation in weak interaction fundamentally may intimately
connected to the magnetic monopoles. The relevance to the $U_A(1)$ problem is
also discussed. The existence of colored magnetic monopoles may also solve the
$U_A(1)$ problem. In the presence of U(1) or monopoles as color singlets, the
't Hooft's solution to the $U_A(1)$ problem is expected. The quantization
formula for the vortex structure is also derived. In the presence of a magnetic
monopole of topological charge $n\neq 0$ in non-abelian gauge theories, the
relevant integral for the vortex along a closed loop in the gauge orbit space
is quantized as $4N\pi/n$ with integer N being the Pontryagin topological
number for the relevant gauge functions.
\end{abstract}
\newpage
\section{Introduction}
Since the discovery of Yang-Mills theories$^2$, particle physics has gained
great development in the frame work of non-abelian gauge theories. One of the
most interesting features of particle physics is the non-perturbative effects
in
gauge theories such as instanton$^3$ effects and magnetic monopoles$^{1,4-5}$.
One of the other most interesting features in non-abelian gauge theories is the
strong CP problem in QCD. It is known that, in non-abelian gauge theories a
Pontryagin or $\theta$ term,
\begin{equation}
{\cal L}_{\theta}=\frac{\theta}{32{\pi}^2}\epsilon^{\mu\nu\lambda\sigma}F^
a_{\mu\nu}F^a_{\lambda\sigma},
\end{equation}
can be added to the Lagrangian density of the system due to instanton effects.
With an arbitrary value of $\theta$, it can induce CP violations. However, the
interesting fact is that the $\theta$ angle in QCD can be only very small
(${\theta\leq}10^{-9}$) or vanishing$^6$. Where in our discussions of QCD,
$\theta$ denotes $\theta+arg(detM)$ effectively with M being the quark mass
matrix, with the effects of electroweak interactions are included. One of the
most interesting approaches to solve the strong CP problem has been the
assumption of an additional Peccei-Quinn $U(1)_{PQ}$ symmetry$^7$. In this
approach, the vacuum angle is ensured vanishing due to the axions$^{8-9}$
introduced. But there has not been observational support$^6$ to the axions
which are needed in this approach. Therefore, it is of fundamental interest to
consider other possible solutions to the strong CP problem.

One of the main purposes of this paper is to generalize the discussions of a
non-perturbative approach proposed$^{5}$ by the author to solve the strong CP
problem and some relevant applications. This is due to its physical
importance as well as the other physical relevance. The section 3 has
some overlapping with our brief note Ref. 5, this is essential for the
completeness due to its intimate connection to the other aspects of the gauge
theories in our discussions. Our approach is to show that the existence of
magnetic monopoles can ensure the quantization of the $\theta$ angle and thus
can provide the solution to the strong CP problem. We will extend the formalism
of Wu and Zee$^{10}$ for discussing the effects of the Pontryagin term in pure
Yang-Mills theories in the gauge orbit spaces in the Schrodinger formulation.
This formalism is useful to the understanding of topological effects in gauge
theories, it has also been used with different methods to derive the mass
parameter quantization in three-dimensional Yang-Mills theory with Chern-Simons
term$^{10-11}$. Wu and Zee showed$^{10}$ that the Pontryagin term induces an
abelian background field in the gauge configuration space of the Yang-Mills
theory. In our discussions, we will consider the case with the existence of a
magnetic monopole. Especially, we will show that magnetic monopoles in the
space will induce an abelian gauge field with non-vanishing field strength in
gauge configuration space, and there can be non-vanishing magnetic flux through
a two-dimensional sphere in the gauge orbit space. Then, the Dirac quantization
conditions$^{4-5}$ in the corresponding quantum theories ensure that the
relevant vacuum angle $\theta$ must be quantized. The quantization rule is
derived as $\theta=2\pi/n$ with n being the topological charge of the monopole
to be given. Therefore, the strong CP problem is automatically solved with the
existence of magnetic monopoles of charge $\pm 1$, or monopoles with very large
total magnetic charges $(n\geq 10^92\pi)$.
As we will see that an interesting feature in our derivation is that the Dirac
quantization condition both in the ordinary space and the relevant induced
gauge orbit space will be used. The relevance to the $U_A(1)$ problem will also
be discussed. We will also discuss about the vortex structure in the gauge
configuration space in this case. As we will show that the vortex in the gauge
orbit space must be quantized also intimately connected to the quantization
rule for the vacuum angle $\theta$. In the presence of a magnetic monopole of
topological charge $n\neq 0$, the relevant integrals for the vortex along a
closed loops in the gauge orbit space are quantized as $4N\pi/n$ with integers
N
being the Pontryagin topological numbers for the relevant gauge functions.

This paper will be organized as follows. Next, we will first give a brief
description of the Schrodinger formulation for our purpose. Then in section 2,
we will clarify the topological results relevant to our discussions. In section
3, we will show the existence of the monopole structure in the relevant gauge
orbit space and realize the relevant topological results explicitly. In section
4, we will discuss about the monopole structure as a solution to the strong CP
problem and its relevance to the $U_A(1)$ problem. The section 5 will be mainly
discussions of the vortex structure in the gauge orbit space in the presence of
a magnetic monopole. Our conclusions will be summarized in section 6.

We will now consider the Yang-Mills theory with the existence of a magnetic
monopole at the origin. The Lagrangian of the system with the $\theta$ term is
given by
\begin{equation}
{\cal L}={\int}d^4x\{-\frac{1}{4}F^a_{\mu\nu}F^{a\mu\nu}+\frac{\theta}{32\pi^2}
\epsilon^{\mu\nu\lambda\sigma}F^a_{\mu\nu}F^{a\lambda\sigma}\}.
\end{equation}
We will choose Weyl gauge $A_0=0$. This is convenient since effectively $A_0$
is not relevant to abelian gauge structure in the gauge configuration space
with the $\theta$ term included. The conjugate momentum corresponding to
$A^a_i$ is then
\begin{equation}
\pi^a_i=\frac{\delta{\cal L}}{\delta\dot{A}^a_i}=\dot{A}^a_i+\frac{\theta}
{8\pi^2}\epsilon_{ijk}F^a_{jk}.
\end{equation}
In the Schrodinger formulation, the system is similar to the quantum system
of a particle with the coordinate $q_i$ moving in a gauge field $A_i(q)$ with
the correspondence$^{10-11}$
\begin{eqnarray}
q_i(t)\rightarrow A^a_i({\bf x},t),\\
A_i(q)\rightarrow {\cal A}^a_i({\bf A}({\bf x})),
\end{eqnarray}
where
\begin{eqnarray}
{\cal A}^a_i({\bf A}({\bf x}))=\frac{\theta}{8\pi^2}\epsilon_{ijk}F^a_{jk}.
\end{eqnarray}
Thus there is a gauge structure with gauge potential ${\cal A}$ in this
formalism within a gauge theory with the ${\theta}$ term included. According to
this, the system can be described by a Hamiltonian equation$^{10}$ or in the
path integral formalism$^{11}$. We will not discuss about this here, since we
only need the Dirac quantization condition for our purpose. For details, see
Ref. 10 and 11.
\section{Relevant Topological Results}
In our discussions, We will use the convention in Ref. 10 and differential
forms$^{12}$ where $A=A^a_iL^adx^i, F=\frac{1}{2}F^a_{jk}L^adx^jdx^k$ with
$F=dA+A^2$, and $tr(L^aL^b)=-\frac{1}{2}{\delta}^{ab}$ in a basis
$\{L^a\mid a=1, 2,...,rank(G)\}$ for the Lie algebra of the gauge group G.
In quantum theory, the Schrodinger formulation is described in the gauge orbit
space with the constraint of Gauss' law. Let ${\cal U}$ denote the gauge
configuration space consisting of all the well-defined gauge potentials $A$
that transform as $A^g=g^{-1}Ag+g^{-1}dg$ under a gauge transformation with
gauge function g. Denoting by ${\cal G}$ the space of all the continuous gauge
transformations, the gauge orbit space ${\cal U}$/${\cal G}$ is the quotient
space of the gauge configuration space with gauge potentials connected by
continuous gauge transformations as equivalent. In the presence of a magnetic
monopole, generally a singularity-free gauge potential may need to be
defined in each local coordinate region. The separate gauge potentials in an
overlapping region can only differ by a continuous gauge transformation$^5$.
In fact, the single-valuedness of the gauge function in the overlapping regions
corresponds to the Dirac quantization condition$^{5}$. For a monopole at the
origin, one can actually divide the space outside the monopole into two
overlapping regions. At a given r, the regions are two extended semi-spheres
around the monopole, with $\theta
\in[\pi/2-\delta,\pi/2+\delta] (0<\delta<\pi/2)$ in the overlapping region,
where the $\theta$ denotes the $\theta$ angle in the spherical polar
coordinates.

As we will see that our equation for our quantization rule for the
$\theta$ is determined by the integration on the space boundary which is
topologically a 2-sphere $S^2$ for non-singular monopoles. Thus for the
quantization of $\theta$, the relevant case is that gauge potentials and gauge
functions are restricted on the space boundary $S^2$. We will call the induced
spaces of ${\cal U}$, ${\cal G}$ and ${\cal U}$/${\cal G}$ with A and g
restricted on the space boundary as restricted gauge configuration space,
restricted space of gauge transformations and the restricted gauge orbit space
respectively. Collectively, they will be called as the restricted spaces,
and the unrestricted ones will be called as usual spaces. We will use the same
notation ${\cal U}$, ${\cal G}$ and ${\cal U}$/${\cal G}$ for both of them for
convenience, there should not be confusing. Our discussions for the monopole
and vortex structures will be on the restricted and usual spaces respectively.

The topological discussions and the results we will now give are true both for
the usual spaces and the restricted spaces. Since ${\cal U}$ is topologically
trivial both for the usual and restricted gauge configuration spaces as we will
see.

To establish the topological results we need, we note first that ${\cal U}$ is
topologically trivial, thus $\Pi_N({\cal U})=0$ for any N. This is due to the
fact that the interpolation between any two gauge potentials $A_1$ and $A_2$
\begin{equation}
A_t=tA_1+(1-t)A_2
\end{equation}
for any real t is also a gauge potential, thus $A_t\in{\cal U}$ (Theorem 7
in Ref.9, and Ref.6). since $A_t$ is transformed as a gauge potential in each
local coordinate region, and in an overlapping region, both $A_1$ and $A_2$ are
gauge potentials may be defined up to a gauge transformation, then $A_t$ is a
gauge potential which may be defined up to a gauge transformation in the
overlapping regions, or $A_t\in{\cal U}$.

The space $\cal U$ can be considered as a bundle over the base space
${\cal U}/{\cal G}$ with fiber ${\cal G}$. More generally for a bundle
$\beta=\{B,P,X,Y,\bar{G}\}$ with bundle space B, base space X, fiber Y, group
$\bar{G}$, and projection P, let $Y_0$ be the fiber over $x_0\in X$, and
let $i:Y_0\rightarrow B$ and $j:B\rightarrow (B,Y_0)$ be the inclusion maps.
Then we have the homotopy sequence$^{13}$ of $(B,Y_0,y_0)$ given by
\begin{equation}
\Pi_N(Y_0)\stackrel{i_*}{\longrightarrow}\Pi_N({B})\stackrel{j_*}
{\longrightarrow}\Pi_N(B,Y_0)\stackrel{\partial_*}{\longrightarrow}
\Pi_{N-1}(Y_0)\stackrel{i_*}{\longrightarrow}\Pi_{N-1}(B)~ (N\geq 1),
\end{equation}
where $\partial$ is the natural boundary operator, $i_*, j_*$ and $\partial_*$
are maps induced by i, j and $\partial$ respectively. Let $P_0$ denote the
restriction of P as a map $(B,Y_0,y_0){\rightarrow}(X,x_0,x_0)$. Then
$P_0j$ is the projection $p:(B,y_0)\rightarrow(X,x_0,B)$. We have the
isomorphism relation
\begin{equation}
p_*:\Pi_N(B,Y_0)\cong\Pi_N(X,x_0).
\end{equation}
Defining $\Delta_*=\partial(P_{0*})^{-1}:\Pi_N(X,x_0)\longrightarrow
\Pi_{N-1}(Y_0,y_0)$, the exact homotopy sequence can be written as
\begin{equation}
\Pi_N(Y_0,y_0)\stackrel{i_*}{\longrightarrow}\Pi_N({B})\stackrel{p_*}
{\longrightarrow}\Pi_N(X,x_0)\stackrel{\Delta_*}{\longrightarrow}
\Pi_{N-1}(Y_0,y_0)\stackrel{i_*}{\longrightarrow}\Pi_{N-1}(B)~ (N\geq 1).
\end{equation}
Now for our purpose with $B={\cal U}, X={\cal U}/{\cal G}$, $Y={\cal G}$,
and $\bar{G}=G$ for the gauge group G. The choice of the base points $x_0$ and
$y_0$ are irrelevant in our discussions, since all the relevant homotopy groups
based on different points are isomorphic. Note that homotopy theory has also
been used to study the global gauge anomalies $^{14-22}$, especially by
using extensively the exact homotopy sequences of fiber bundles and in terms of
James numbers of Stiefel manifolds.

More explicitly, we can now consider the following exact homotopy
sequence$^{13}$:
\begin{equation}
\Pi_N({\cal U})\stackrel{P_*}{\longrightarrow}\Pi_N({\cal U}/{\cal G})
\stackrel{\Delta_*}{\longrightarrow}\Pi_{N-1}({\cal G})\stackrel{i_*}
{\longrightarrow}\Pi_{N-1}({\cal U}) ~ (N\geq 1).
\end{equation}
Since as we have seen that $\Pi_N({\cal U})=0$ for any N, we have
\begin{equation}
0\stackrel{P_*}{\longrightarrow}\Pi_N({\cal U}/{\cal G})
\stackrel{\Delta_*}{\longrightarrow}\Pi_{N-1}({\cal G})\stackrel{i_*}
{\longrightarrow}0 ~ (N\geq 1).
\end{equation}
This implies that
\begin{equation}
\Pi_N({\cal U}/{\cal G})\cong\Pi_{N-1}({\cal G})~ (N\geq 1).
\end{equation}
As shown by Wu and Zee for the usual spaces in pure Yang-Mills theory in four
dimensions,
\begin{equation}
\Pi_1({\cal U}/{\cal G})\cong\Pi_{0}({\cal G})
\end{equation}
is non-trivial, and thus $\theta$ term induces a vortex structure in gauge
orbit
space. This isomorphism will also be used in our discussions of the vortex
structure in the presence of a magnetic monopole, but as we will see that it's
explicit realization is more non-trivial. It was also showed in Ref. 10 that
the field strength $\cal F$ associated with the gauge potential $\cal A$ is
vanishing, and thus there is no flux corresponding to $\cal F$ in the pure
Yang-Mills theory.

However, as we will show in the next section that in the presence of a magnetic
monopole, the relevant topological properties of the system are drastically
different. This will give interesting consequences in the quantum theory. One
of the main topological result we will use for the restricted spaces in the
presence of a magnetic monopole is
\begin{equation}
\Pi_2({\cal U}/{\cal G})\cong\Pi_{1}({\cal G}).
\end{equation}

Now $\Pi_2({\cal U}/{\cal G})\neq 0$ corresponds to the condition for the
existence of a magnetic monopole in the restricted gauge orbit space. In the
next section, we will realize the above topological results. We will first show
that in this case ${\cal F}\neq 0$, and then demonstrate explicitly that the
magnetic flux $\int_{S^2}\hat{\cal F}\neq 0$ can be nonvanishing in the
restricted gauge orbit space, where $\hat{\cal F}$ denotes the projection of
$\cal F$ into the restricted gauge orbit space.
\section{Monopole Structure in the Restricted Gauge Orbit Space in the Presence
of Magnetic Monopoles}
In our discussions, we denote the differentiation with respect to space
variable ${\bf x}$ by d, and the differentiation with respect to
parameters $\{t_i\mid i=1,2...\}$ which {\bf A}({\bf x}) may depend on in the
gauge configuration space by $\delta$, and assume $d\delta+\delta d$=0. Then,
similar to $A=A_{\mu}dx^{\mu}$ with $\mu$ replaced by a, i, ${\bf x}$, the
gauge potential in the gauge configuration space can be written as a 1-form
given by
\begin{equation}
{\cal A}=\int d^3x{\cal A}^a_i({\bf A}({\bf x}))\delta A^a_i(\bf x).
\end{equation}
Using Eq.(6), this gives
\begin{equation}
{\cal A}=\frac{\theta}{8\pi^2}\int d^3x\epsilon_{ijk}F^a_{jk}({\bf x})\delta
A^a_i({\bf x})=-\frac{\theta}{2\pi^2}\int_M tr(\delta AF),
\end{equation}
with M being the space manifold. Since ${\cal A}$ is an abelian, then the field
strength is given by
\begin{equation}
{\cal F}=\delta{\cal A}.
\end{equation}
With $\delta F=-D_A(\delta A)=-\{d(\delta A)+A\delta A-\delta AA\}$, we have
\begin{equation}
{\cal F}=\frac {\theta}{2\pi^2}\int_Mtr[\delta AD_A(\delta A)]
=\frac {\theta}{4\pi^2}\int_Mdtr(\delta A\delta A)
=\frac {\theta}{4\pi^2}\int_{\partial M}tr(\delta A\delta A),
\end{equation}
up to a local term with vanishing projection to the relevant gauge orbit space.
Usually, one may assume $A\rightarrow 0$ faster than 1/r as $\bf x\rightarrow
0$
, then this would give ${\cal F}=0$ as in the case of pure Yang-Mills
theory$^{10}$. However, it is more subtle in the presence of a magnetic
monopole. Asymptotically as $r\rightarrow 0$ with a monopole at the origin, the
monopole may generally give a field strength of the form$^{4-5,22}$
\begin{equation}
F_{ij}=\frac{1}{4\pi r^2}\epsilon_{ijk}({\bf {\hat r}})_kG_0({{\bf \hat r}}),
\end{equation}
with $\bf {\hat r}$ being the unit vector for {\bf r}, and this gives
$A\rightarrow O(1/r)$ as $\bf x\rightarrow 0$. Thus, one can see
easily that a magnetic monopole can give a nonvanishing field strength $\cal F$
in the gauge configuration space.

To evaluate $\cal F$, one needs to specify the space boundary $\partial M$ in
the presence of a magnetic monopole. We now consider the case that the magnetic
monopole does not generate a singularity in the space. Then the effects in the
case that monopoles are singular will be discussed. In fact, non-singular
monopoles may be more relevant in the unification theory since there can be
monopoles as a smooth solution of a spontaneously broken gauge theory similar
to 't Hooft Polyakov monopole$^4$. For example, it is known that$^{23}$ there
are monopole solutions in the minimal SU(5) model. When the monopole is
non-singular, the space boundary then may be regarded as a large 2-sphere $S^2$
at the spatial infinity. For our purpose, we actually only need to evaluate the
projection of $\cal F$ into the gauge orbit space. But the evaluation of
$\cal F$ can give more explicit understanding of the topological properties of
the system. The $\cal F$ is similar to a constant F in the ordinary space, it
does not give any flux through a closed surface in the space $\cal U$. However,
the quantum theory is based on the gauge orbit space in Schrodinger formulation
, the relevant magnetic flux needs to considered in the gauge orbit space. In
fact, as we will see that the corresponding magnetic flux in the gauge orbit
space can be non-vanishing. A gauge potential in the gauge orbit space can be
written in the form of
\begin{equation}
A=g^{-1}ag+g^{-1}dg,
\end{equation}
for an element a $\in{\cal U}/{\cal G}$ and a gauge function $g\in{\cal G}$.
Then the projection of a form into the gauge orbit space contains only terms
proportional to $(\delta a)^n$ for integers n. We can now write
\begin{equation}
\delta A=g^{-1}[\delta a-D_a(\delta gg^{-1})]g.
\end{equation}
Then we obtain
\begin{equation}
{\cal A}=-\frac{\theta}{2\pi^2}\int_M tr(f\delta a)
+\frac{\theta}{2\pi^2}\int_M tr[fD_a(\delta gg^{-1})],
\end{equation}
where $f=da+a^2$. With some calculations, this can be simplified as
\begin{equation}
{\cal A}=\hat{\cal A}
+\frac{\theta}{2\pi^2}\int_{S^2}tr[f\delta gg^{-1}],
\end{equation}
where
\begin{equation}
\hat{\cal A}=-\frac{\theta}{2\pi^2}\int_M tr(f\delta a),
\end{equation}
is the projection of $\cal A$ into the gauge orbit space. Similarly, we have
\begin{equation}
{\cal F}=\frac {\theta}{4\pi^2}\int_{S^2}
tr\{[\delta a-D_a(\delta gg^{-1})][\delta a-D_a(\delta gg^{-1})]\}
\end{equation}
or
\begin{equation}
{\cal F}=\hat{\cal F}-\frac {\theta}{4\pi^2}\int_{S^2}
tr\{\delta aD_a(\delta gg^{-1})+D_a(\delta gg^{-1})\delta a
-D_a(\delta gg^{-1})D_a(\delta gg^{-1})\},
\end{equation}
where
\begin{equation}
\hat{\cal F}=\frac {\theta}{4\pi^2}\int_{S^2}tr(\delta a\delta a),
\end{equation}
is the projection of the ${\cal F}$ to the gauge orbit space or the restricted
gauge orbit space based on the space boundary $S^2$.

Now all our discussions will be based on the restricted spaces.
To see that the flux of $\hat{\cal F}$ through a closed surface in
the gauge orbit space ${\cal U}/{\cal G}$ can be nonzero, we will construct
a 2-sphere in it. Consider an given element $a\in{\cal U}/{\cal G}$, and a loop
in $\cal G$. The set of all the gauge potentials obtained by all the gauge
transformations on $a$ with gauge functions on the loop then forms a loop
$C^1$ in the gauge configuration space $\cal U$. Obviously, the $a$ is the
projection of the loop $C^1$ into ${\cal U}/{\cal G}$.
Now since $\Pi_1({\cal U})=0$ is trivial, the loop $C^1$ can be continuously
extended to a two-dimensional disc $D^2$ in the $\cal U$ with the boundary
$\partial D^2=C^1$. Obviously, the projection of the $D^2$ into the gauge orbit
space with the boundary $C^1$ identified as a single point is topologically a
2-sphere $S^2\subset{\cal U}/{\cal G}$. With the Stokes' theorem in the gauge
configuration space, We now have
\begin{equation}
\int_{D^2}{\cal F}=\int_{D^2}\delta{\cal A}=\int_{C^1}{\cal A}.
\end{equation}
Using Eqs.(24) and (29) with $\delta a=0$ on $C^1$, this gives
\begin{equation}
\int_{D^2}{\cal F}=\int_{C^1}{\cal A}
=\frac{\theta}{2\pi^2}tr\int_{S^2}\int_{C^1}[f\delta gg^{-1}].
\end{equation}
Thus, the projection of the Eq(30) to the gauge orbit space gives
\begin{equation}
\int_{S^2}\hat{\cal F}
=\frac{\theta}{2\pi^2}tr\int_{S^2}\{f\int_{C^1}\delta gg^{-1}\},
\end{equation}
where note that in the two $S^2$ are in the restricted gauge orbit space and
the ordinary space respectively. This can also be obtained by
\begin{equation}
\int_{D^2}tr\int_{S^2}
tr\{\delta aD_a(\delta gg^{-1})+D_a(\delta gg^{-1})\delta a
-D_a(\delta gg^{-1})D_a(\delta gg^{-1})\}=0,
\end{equation}
or the projection of $\int_{D^2}{\cal F}$ gives $\int_{S^2}\hat{\cal F}$. We
have verified this explicitly or the topological result that the projection of
$\int_{D^2}{\cal F}$ gives $\int_{S^2}\hat{\cal F}$. For this one needs to use
Stokes theorem in the ordinary space and the gauge configuration space with
$d\delta+\delta d=0$, $a\in{\cal U}/{\cal G}$ or $a$ is a constant on $C^1$,
and $\int_{D^2}\hat{\cal F}=\int_{S^2}\hat{\cal F}$ in the gauge orbit space
since $\hat{\cal F}$ is the projection of the ${\cal F}$ into the gauge orbit
space.

In quantum theory, Eq.(31) corresponds to the topological result
$\Pi_2({\cal U}/{\cal G})\cong\Pi_{1}({\cal G})$ for the restricted spaces.
This feature in the gauge orbit space has some similarity to that given in
Refs.10 and 11 for the discussions of three-dimensional Yang-Mills theories
with a Chern-Simons term. We only need the Dirac quantization condition here
for our purpose. In the restricted gauge orbit space, the Dirac quantization
condition gives
\begin{equation}
\int_{S^2}\hat{\cal F}=2\pi k,
\end{equation}
with k being integers. We will now determine the quantization rule for the
$\theta$. Now let $f$ be the field strength 2-form for the magnetic monopole.
There may be many ways to obtain non-vanishing results for the right-hand side.
For our purpose, one way is to restrict $g$ to a $U(1)$ subgroup of the gauge
group, and obtain a non-zero topological number. Then the quantization rule for
the $\theta$ will be obtained.

Let $\{H_i\mid i=1, 2,...,r=rank(G)\}$ denote a basis of the Cartan subalgebra
for the gauge group G. The corresponding simple roots and fundamental weights
are denoted by $\{\alpha_i\mid i=1, 2,...,r\}$ and
$\{\lambda_i\mid i=1, 2,...,r\}$ respectively. Then we have$^{24}$
\begin{equation}
\frac{2<\lambda_i,\alpha_j>}{<\alpha_j,\alpha_j>}=\delta_{ij},
\end{equation}
where the $<\lambda_i,\alpha_j>$ denotes the inner product in the root vector
space. By the theorem which states that for any compact and connected Lie group
G
, any element in the Lie algebra is conjugate to at least one element in its
Cartan subalgebra by a group element in G, the quantization condition for the
magnetic monopole is given by$^{23}$
\begin{equation}
exp\{\int_{S^2}f\}=exp\{G_0\}=exp\{4\pi\sum_{i=1}^{r}\beta^{i}H_{i}\}\in Z.
\end{equation}
Where
\begin{equation}
G_0=\int_{S^2}f=4\pi\sum_{i=1}^{r}\beta^{i}H_{i}
\end{equation}
is the magnetic charge up to a conjugate transformation by a group element.

Now let g(t) $t\in [0,1]$ be in the following U(1) subgroup on the $C^1$
\begin{equation}
g(t)=exp\{4\pi mt\sum_{i,j=1}^{r}\frac{(\alpha_i)^jH_j}{<\alpha_i,\alpha_i>}\},
\end{equation}
with m being integers. In fact, m should be identical to k according to our
topological result $\Pi_2({\cal U}/{\cal G})\cong\Pi_{1}({\cal G})$ for the
restricted spaces. In this case, the relevant homotopy groups obtained are
isomorphic to Z which may be only a subgroup of the homotopy groups generally
for a non-abelian gauge group G. In fact for this case, the k and m should be
identical since they correspond to the topological numbers on each side. Using
$tr(H_iH_j)=-\frac{1}{2}\delta_{ij}$ and
\begin{equation}
\int_{C^1}\delta gg^{-1}
=4\pi m\sum_{i,j=1}^{r}\frac{(\alpha_i)^jH_j}{<\alpha_i,\alpha_i>},
\end{equation}
we obtain
\begin{equation}
\theta=\frac{2\pi}{n}~(n\neq0).
\end{equation}
Where we define generally the topological charge of the magnetic monopole as
\begin{equation}
n=-2<\delta,\beta>=-2\sum_{i=1}^{r}<\lambda_i,\beta>,
\end{equation}
which must be an integer by the quantization condition$^{23}$ for the magnetic
monopoles. Where the $\delta$ is given by
\begin{equation}
\delta=\sum_{i=1}^{r}\frac{2\alpha_i}{<\alpha_i,\alpha_i>}
=\sum_{i=1}^{r}\lambda_i.
\end{equation}
The minus sign is due to our normalization convention for the Lie algebra
generators. Actually, the fundamental weights $\{\lambda_i\mid i=1, 2,...,r\}$
and $\{\frac{2\alpha_i}{<\alpha_i,\alpha_i>}\mid i=1, 2,...,r\}$ form the
Dynkin
basis and its dual basis in the root vector space respectively.

In our definition, the topological charge of the magnetic monopole can be
understood as follows. Up to a conjugate transformation, the magnetic charge
of the monopole is contained in a Cartan subalgebra of the gauge group.
Restricting to each U(1) subgroup generated by a generator $H_i$ (i=1, 2,...,r)
in the basis of the Cartan subalgebra, the monopole has a topological number
$n_i$ corresponding to the Dirac quantization condition. Then generally the
topological number n in our definition is given by
\begin{equation}
n=\sum_{i=1}^{r}n_{i}.
\end{equation}
Obviously, we expect that this is the natural generalization of the topological
charge to the non-abelian magnetic monopole. To the knowledge of the author,
such an explicit general definition Eq.(40) in terms of the fundamental weights
of the Lie algebra for the topological charge of non-abelian magnetic monopoles
is first obtained by the author.

As a remark, our derivation has been topological. Our quantization rule can
also be obtained by using constraints of Gauss' law. This more physical
approach and the discussion of its physical relevance will be given elsewhere.
\section{Magnetic Monopoles as A Solution to the Strong CP Problem and the
Relevance to the $U_A(1)$ Problem}
As we have seen that in the presence of magnetic monopoles, the vacuum angle
$\theta$ must be quantized. The quantization rule is given by Eq.(39).
Therefore, we conclude that the existence of magnetic monopoles can provide a
solution to the strong CP problem. In the presence of magnetic monopoles with
topological charge $\pm 1$, the vacuum angle of non-abelian gauge theories
must be $\pm 2\pi$, the existence of such magnetic monopoles gives a solution
to the strong CP problem. The existence of many monopoles can ensure
$\theta\rightarrow 0$, and the strong CP problem may also be solved. In this
possible solution to the strong CP problem with $\theta\leq 10^{-9}$, the total
magnetic charges present are $|n|\geq 2\pi 10^{9}$. This may possibly be within
the abundance allowed by the ratio of monopoles to the entropy$^{26}$, but with
the possible existence of both monopoles and anti-monopoles, the total number
of magnetic monopoles may be larger than the total magnetic charges. Generally,
one needs to ensure that the total number is consistent with the experimental
results on the abundance of monopoles.

In the above discussions, we consider the case that magnetic monopole generates
no singularity in the space, for example, with monopole as a smooth solution
in a spontaneously broken gauge theory. If we consider the magnetic monopoles
as
a singularity similar to the Wu-Yang monopole$^{27}$ which is the first
non-abelian monopole solution found, then the space boundary can be regarded as
consisting of an infinitesimal inward 2-sphere around each magnetic monopole
and a large 2-sphere at the spatial infinity. In the space outside the monopole
, each infinitesimal inward sphere effectively gives a contribution equivalent
to a monopole of opposite topological charge. Then the total contribution of
the infinitesimal spheres and the contribution from the large sphere at the
spatial boundary are all cancelled in the relevant integrations. Therefore,
only the existence of non-singular magnetic monopoles may provide solution to
the strong CP problem.

Moreover, note that our conclusions are also true if we add an additional
$\theta$ term in QED with the $\theta$ angle the same as the effective
$\theta$ in QCD if there exist Dirac monopoles as color singlets, or a
non-abelian monopoles with magnetic charges both in the color SU(3) and
electromagnetic U(1). Then the explanation of such a QED $\theta$ term is
needed. The effect of the term proportional to $\epsilon^{\mu\nu\lambda\sigma}
F_{\mu\nu}F_{\lambda\sigma}$ in the presence of magnetic charges was first
considered$^{28,29}$ relevant to chiral symmetry. Then, the effect of a similar
U(1) $\theta$ term was discussed for the purpose of considering the induced
electric charges$^{30}$ as quantum excitations of dyons associated with the
't Hooft Polyakov monopole and generalized magnetic monopoles$^{23,31}$.
Especially, the generalized magnetic monopoles are used to consider the
possibility of quarks as dyons in a spontaneously broken gauge theory$^{31}$.
An
interesting feature is that$^{31}$ if quarks are dyons in a spontaneously
broken
gauge theory, then their electric charges will not be exactly fractionally
quantized, instead they will carry extra charges proportional to $\theta$.
Moreover, two meson octets, one baryon octet and one baryon decuplet free of
magnetic charges were constructed$^{31}$ from quarks as dyons. For our purpose,
we expect that if a QED $\theta$ term is included, it may be possibly an
indication of unification for the color gauge symmetry and electromagnetic U(1)
symmetry. A $\theta$ term needs to be included in the unification gauge theory
since $\Pi_3(G)=Z$ for the unification group G, monopoles with magnetic charges
involving the QED U(1) symmetry are generated through spontaneous gauge
symmetry breaking. Generally, such an arbitrary $\theta$ term in QED may not be
discarded since it is not a total divergence globally in the presence of
magnetic charges and as we have seen that the $\theta$ term physically can have
non-perturbative effects.

We would like to emphasize that in our approach with magnetic monopoles as a
solution to the strong CP problem, the $U_A(1)$ problem can also be solved. The
$U_A(1)$ problem was originally solved by t' Hooft$^{3}$ with the fact that
the $U_A(1)$ is not a symmetry in the quantum theory due to the axial
anomaly$^{32}$, the conserved $U_A(1)$ symmetry is not gauge invariant and its
spontaneous breaking does not generate a physical light meson. In our solution
with the presence of colored magnetic monopoles, the $U_A(1)$ symmetry is
explicitly broken$^{28}$. If the strong CP problem is solved by the pure
electromagnetic U(1) monopole with a $\theta$ term included with $\theta$ being
the same as the effective $\theta_{QCD}$, then the $U_A(1)$ symmetry is not
explicitly broken and the t' Hooft's solution to the $U_A(1)$ problem can be
applied. Thus $U_A(1)$ problem can be solved in our approach to solve the
strong
CP problem with the existence of magnetic monopoles.

\section{Vortex Structure in the Gauge Orbit Space}
In this section, we will discuss about the vortex structure in the gauge
orbit space. It is known that$^{33}$ there can be vortex structures in some
three-dimensional field theories with the boundary of the space being
topologically a circle. The discussions in Ref.10 in the gauge orbit space are
for the pure Yang-Mills theory with a $\theta$ term. We will consider the case
in the presence of magnetic monopoles.

In order to discuss about the vortex structure in the gauge orbit space in the
gauge theories we are interested in, we need to consider the integration of
$\hat{\cal A}$ along a closed loop $\hat C$ in the gauge orbit space
${\cal U}/{\cal G}$. As in Ref.10, such a loop $\hat C$ can be constructed by
projection. Let C denote an open path in the gauge configuration space
${\cal U}$ with gauge potentials $A$ and $A^g$ as the two end points, where
$g\in{\cal G}$ is a gauge function. Obviously, the projection of C into the
${\cal U}/{\cal G}$ gives a closed loop $\hat{C}$ with the two end points of C
identified as a single point. Thus we have$^{10}$ topologically
\begin{equation}
\int_{\hat C}\hat{\cal A}\cong\int_C{\cal A}.
\end{equation}
In pure Yang-Mills theory, one can verify that$^{10}$ the ${\cal A}$ can be
written as the differentiation of the Chern-Simons secondary topological
invariant$^{34}$ in the gauge configuration space. Thus in the case of pure
Yang-Mills theory, the Chern-Simons secondary invariant can be regarded as a
gauge function in the gauge configuration space. Now in the presence of a
magnetic monopole, we obtain
\begin{equation}
\delta\{\theta W[A]\}=-\frac{\theta}{2\pi^2}\int_M\tr(\delta AF)
+\frac{\theta}{4\pi^2}\int_M dtr(A\delta A),
\end{equation}
or
\begin{equation}
{\cal A}=\delta\{\theta W[A]\}-\frac{\theta}{4\pi^2}\int_M dtr(A\delta A),
\end{equation}
where
\begin{equation}
W[A]=-\frac{1}{4\pi^2}\int_Mtr(AdA+\frac{2}{3}A^3)
\end{equation}
is the Chern-Simons secondary topological invariant.
This gives
\begin{equation}
\int_C{\cal A}=\int_C\delta\{\theta W[A]\}
-\frac{\theta}{4\pi^2}\int_C\int_{\partial M}dtr(A\delta A),
\end{equation}
or
\begin{equation}
\int_C{\cal A}=\theta\{W[A^g]-W[A]\}
-\frac{\theta}{4\pi^2}\int_C\int_{S^2}tr(A\delta A).
\end{equation}
Now
\begin{equation}
W[A^g]-W[A]=\frac{1}{12\pi^2}\int_Mtr(g^{-1}dg)^3+\int_Md\alpha_2[A,g]
=2N[g]+\int_{S^2}\alpha_2[A,g],
\end{equation}
where
\begin{equation}
N[g]=\frac{1}{24\pi^2}\int_Mtr(g^{-1}dg)^3,
\end{equation}
and
\begin{equation}
\alpha_2[A,g]=-\frac{1}{4\pi^2}tr(Ag^{-1}dg).
\end{equation}
Thus
\begin{equation}
\int_C{\cal A}=2\theta N[g]
-\frac{\theta}{4\pi^2}\int_{S^2}tr(Ag^{-1}dg)
-\frac{\theta}{4\pi^2}\int_C\int_{S^2}tr\{A\delta A\}.
\end{equation}
Now since C is an open path in ${\cal U}$ from $A$ to $A^g$, the integral
$\int_C{\cal A}$ generally contains two parts. The first part is topologically
invariant as we will see, the second term depends only on the end points $A$
and
$A^g$ or the gauge function g. The third therm or the second and third term
together is generally path dependent, but it does not contain any non-vanishing
topological invariant, namely it is a path-dependent local term. This can be
seen as follows. Since the space ${\cal U}$ is topologically trivial, the open
path C in ${\cal U}$ with the two end points fixed can be continuously deformed
into the straight interval
\begin{equation}
A_t=tA^g+(1-t)A~~~ t\in [0,1].
\end{equation}
We only need to verify this by evaluating the integral with the C being the
straight interval. Then topologically
\begin{eqnarray}
\int_Ctr\{A\delta A\}{\cong}-\int^1_0A_t(A^g-A)dt
=-tr\{(A^g-A)\int^1_0[(tA^g+(1-t)A)dt\} \nonumber \\
=-tr\{\frac{1}{2}(A^g-A)(A^g+A)\}=-tr[(A^g)^2-2AA^g-A^2]=2tr(AA^g).
\end{eqnarray}
With $A^g=g^{-1}Ag+g^{-1}dg$, this can be written as
\begin{equation}
\int_Ctr\{A\delta A\}\cong 2tr\{Ag^{-1}Ag+Ag^{-1}dg\}.
\end{equation}
Thus topologically
\begin{equation}
\int_C{\cal A}\cong 2\theta N[g]+I_2,
\end{equation}
where
\begin{equation}
I_2=-\frac{\theta}{4\pi^2}\int_{S^2}tr\{(2Ag^{-1}Ag+3Ag^{-1}dg\}.
\end{equation}
One can now easily see that the second part $I_2$ of the integral contains no
non-vanishing topological invariant. Since A and g are independent each other,
the g as a mapping $g:S^2\rightarrow G$ can be continuously deformed into a
constant mapping. The $I_2$ is topologically equivalent to an integral with the
integrand proportional to $trA^2=0$. Thus, up to topologically trivial terms,
we obtain
\begin{equation}
\int_{\hat C}\hat{\cal A}\cong 2\theta N[g].
\end{equation}
It is known that the integral N[g] is topologically invariant when M is
compactified as a 3-sphere $S^3$. It is straightforward to show that N[g] is
topologically invariant with g as a mapping $g:M\rightarrow G$ from the space
manifold to the gauge group G in our discussion. The only change is that a
small variation for the gauge function gives an additional boundary term which
is vanishing due to the fact that the space boundary $\partial M$ is
topologically a 2-sphere and $\Pi_2(G)=0$. To obtain non-vanishing results for
N[g], we need to restrict to the gauge functions with $g\rightarrow 0$. Then
for the space manifold is effectively compactified as a 3-sphere $S^3$, and
N[g]
is the Pontryagin topological number corresponding to the homotopy group
$\Pi_3(G)$. Thus
\begin{equation}
\int_{\hat C}\hat{\cal A}\cong 2\theta N,
\end{equation}
with N being integers. This corresponds to the isomorphism relation
\begin{equation}
\Pi_1({\cal U}/{\cal G})\cong\Pi_{0}({\cal G})=\Pi_3(G)=Z,
\end{equation}
where ${\cal G}$ is the space of all the gauge transformations g satisfying
$g\rightarrow 0$.
With the quantization rule $\theta=2\pi/n$ we obtained, we can write
\begin{equation}
\int_{\hat C}\hat{\cal A}\cong\frac{4N\pi}{n}~~~(n\neq 0),
\end{equation}
with n being the topological charge of the monopole.
Therefore, there can be vortex structure in the gauge orbit space. In the
presence of magnetic monopoles the vortex must be topologically quantized with
the quantization rule given by the above equation. In the presence of a
monopole
of topological charges $\pm 1$, the vortex is quantized as $\pm 4\pi N$. In the
presence of many monopoles with very large total topological charges n, the
vortex can be only very small or vanishing.  Our discussions in the presence of
magnetic monopoles are more non-trivial than the case in pure Yang-Mills theory
, especially for the explicit realization of the topological isomorphism due
to the local terms involved.

In the above discussions, the magnetic monopoles are regarded non-singular in
the space. In fact, one can easily see that in the presence of singular
monopoles the quantization rule is given by Eq.(59), but as we have seen that
in this case $\theta$ can be arbitrary.

As a remark, note that in QED or more generally an abelian gauge theory with
$N=0$ since $\Pi_3(U(1))=0$, there is no corresponding topological vortex in
the gauge orbit space even in the presence of magnetic monopoles.
\section{Conclusions}
We have discussed extensively about the topological structure in the relevant
gauge orbit space of gauge theories with a $\theta$ term. The presence of a
magnetic monopole in the ordinary space can induce monopole and vortex
structures in the restricted and usual gauge orbit spaces. The Dirac
quantization conditions ensure that the vacuum angle $\theta$ must be
quantized. The quantization can provide a solution to the strong CP problem
with the existence of one monopole of topological charge $\pm 1$, or many
monopoles if it is consistent with the abundance of magnetic monopoles. The
$U_A(1)$ problem may also be solved with the existence of colored magnetic
monopoles, or by t' Hooft's solution if the magnetic monopoles are of only
U(1) charges as color singlets. Therefore, the fact that the strong
CP-violation can be only so small or vanishing may be a signal for the
existence of magnetic monopoles. An interesting feature is that
in the presence of one magnetic monopole of charge $\pm 1$, $\theta=\pm 2\pi$
according to the quantization rule obtained. The cases of $n=\pm 2$ may also
possibly solve strong CP problem. But when the vacuum angle is $\theta=\pm\pi$,
other than the Strong CP problem it may have other effects$^{35}$ different
from the case of $\theta=\pm 2\pi$ or vanishing, for example, on quark masses,
but these are usually discussed without the presence of monopoles.
In the axion approach to solve the strong CP problem, the vacuum angle should
be vanishing, and there has been argument$^{35}$ that the vacuum energy is
minimized at vanishing vacuum angle.

We have also derived the quantization formula for the vortex by using our
quantization rule for the $\theta$ angle. Thus, as we have shown that the
monopole structure and vortex structure in the restricted gauge orbit spaces
and the usual gauge orbit spaces are connected through our quantization rules.

As a remark, note that usually if strong interaction conserves CP,
then the $2\pi$ or $\pi$ may be expected equivalent to $-2\pi$ or $-\pi$ since
they are related by a CP operation and $\theta$ may be expected to be periodic
with period $2\pi$. However, according to our quantization rule, this is not
true due to the fact that $\pm 2\pi$ or $\pm\pi$ correspond to different
monopole sectors. If the strong CP problem is solved by a monopole of
topological charge $\pm 1$ or $\pm 2$, this means the CP-violation can be only
very small in the measurements implemented so far, the CP can not be exactly
conserved, since the $\theta=\pm 2\pi$ or $\theta=\pm\pi$ correspond to two
different physical systems. If the strong CP problem is solved due to the
existence of many monopoles, then the observation of strong CP violation gives
an indirect measurement of the abundance of magnetic monopoles. For any finite
number of magnetic monopoles, the CP cannot be exactly conserved in the strong
interactions. The strong CP-violation may provide information about the
structure of the universe.

As a conjecture, we expect that the parity violation and CP violation in weak
interaction may intimately connected to magnetic monopoles also.

The author would like to express his gratitude to Y. S. Wu and A. Zee for
valuable discussions and suggestions. The author is also grateful to O. Alvarez
for his invitation.

\end{document}